\documentclass[twocolumn, article, floatfix, a4paper]{article}

\usepackage{graphicx}

\usepackage{amsmath,amssymb,amsfonts}
\usepackage{color}
\usepackage{hyperref}
\usepackage{textcomp}
\usepackage{verbatim}
\usepackage{epstopdf}
\title{Comparison between Atomic Force Microscopy and Force Feedback Microscopy static force curves} 

\author{\normalsize{Luca Costa$^{1,2}$, Mario S. Rodrigues$^3$, Simon Carpentier$^4$, Pieter Jan van Zwol$^5$, Jo\"{e}l Chevrier$^{1,2,4}$, Fabio Comin$^1$} \\
\begin{minipage}{.80 \linewidth}
\small
$^{1)}$ \textit{European Synchrotron Radiation Facility, 6 rue Jules Horowitz BP 220, 38043 Grenoble Cedex, France} \\
$^{2)}$ \textit{Universit\'{e} Joseph Fourier BP 53, 38041 Grenoble Cedex 9, France} \\
$^{3)}$ \textit{CFMC/Dep. Fısica, Faculdade de Ciências, Universidade de Lisboa, Campo Grande, 1749-016 Lisboa, Portugal} 	\\
$^{4)}$  \textit{Institut N\'{e}el CNRS BP 166, 38042 Grenoble Cedex 9, France} \\ 
$^{5)}$ \textit{ASML, De Run 6501, 5504 DR, Veldhoven, The Netherlands}\\
\end{minipage} \\
\begin{minipage}{.85 \linewidth}
\small
Atomic Force Microscopy (AFM) conventional static force curves and Force Feedback Microscopy (FFM) force curves acquired with the same cantilever at the solid/air and solid/liquid interfaces are here compared. The capability of the FFM to avoid the jump to contact leads to the complete and direct measurement of the interaction force curve, including the attractive short-range van der Waals and chemical contributions. Attractive force gradients five times higher than the lever stiffness do not affect the stability of the FFM static feedback loop. The feedback loop keeps the total force acting on the AFM tip equal to zero, allowing the use of soft cantilevers as force transducers to increase the instrumental sensitivity.
The attractive interactions due to the nucleation of a capillary bridge at the native oxide silicon/air interface or due to a DLVO interaction at the mica/deionized water interface have been measured. This set up, suitable for measuring directly and quantitatively interfacial forces, can be exported to a SFA (Surface Force Apparatus).
\end{minipage}
}

\date{\today}

\begin{document}
\maketitle

\section{\label{sec:level1}Introduction:}
The so called "jump to contact" occurring when the attractive force gradient exceeds the cantilever stiffness characterizes the AFM static force curves \cite{wu}. As a consequence, short-range attractive forces are usually not accessible in conventional static mode.
The increase of the cantilever stiffness is not a convenient solution because it leads to a consistent decrease of the force sensitivity. The complete interaction force curve is then usually obtained with \textit{a posteriori} mathematical data treatment from frequency shifts measured in Frequency Modulation AFM (FM-AFM) \cite{giessibl03, sader04} where stiff cantilevers with high Q-factor are employed.
However, the methods proposed to convert the measured frequency-shift into force gradients require the tip-sample interaction to be conservative over an oscillation cycle of the tip. Interactions including non conservative forces as in the case of capillary condensation \cite{riedo05} at the solid/air interface cannot therefore be precisely measured in FM-AFM. Moreover, the integrated force gradient measured with dynamic AFM techniques is not equivalent to the static force when the interaction has a frequency-dependence behavior as in the case of the mechanical response of viscoelastic materials \cite{navajas03}. Analogous problems in the force reconstruction could come from the different thermodynamic conditions between the static and the dynamic measurements \cite{barcons12}. These remarks emphasize that it is of importance to have a direct and quantitative measurement of the full force curve between two surfaces at nanoscale.\\
An alternative and successful option to dynamic AFM techniques is the use of a magnetic feedback control acting on soft cantilevers aiming to counteract the tip-sample interaction and therefore avoiding the mechanical instability of the tip \cite{jarvis_96,yamamoto97,ashby00}. 
Here a magnetic particle has to be attached on the backside of the cantilevers. The magnetic force is then applied on the tip using a coil positioned below the sample.
Alternative systems commonly labeled by \textit{displacement-controlled scanning force microscopy} \cite{goertz10} have been suggested but the attractive part of the tip-sample interaction has not been sufficiently investigated \cite{kim12,bonander08} as in the case of the magnetic force feedback.\\
A more flexible and simple operational scheme named Force Feedback Microscope (FFM) that doesn't require any special cantilever preparation has recently been proposed \cite{io12}.
It permits to measure simultaneously the force, the conservative force gradient and the damping coefficient, fully characterizing the tip-sample interaction employing soft cantilevers.
The FFM static loop ensures the position of the AFM tip to be stable by imposing a counteracting force on the tip. The total force acting on the tip is therefore maintaining constantly equal to zero.
Here we focus on the measurement of the static force and we compare the ability to measure attractive interactions with a conventional AFM and a FFM.
The measurements have been performed with the very same cantilevers, simply switching between the FFM operational mode into the standard AFM static mode.
\section{\label{sec:level2}Operational scheme:}
\begin{figure}
\begin{center}
\includegraphics[width=8.5cm]{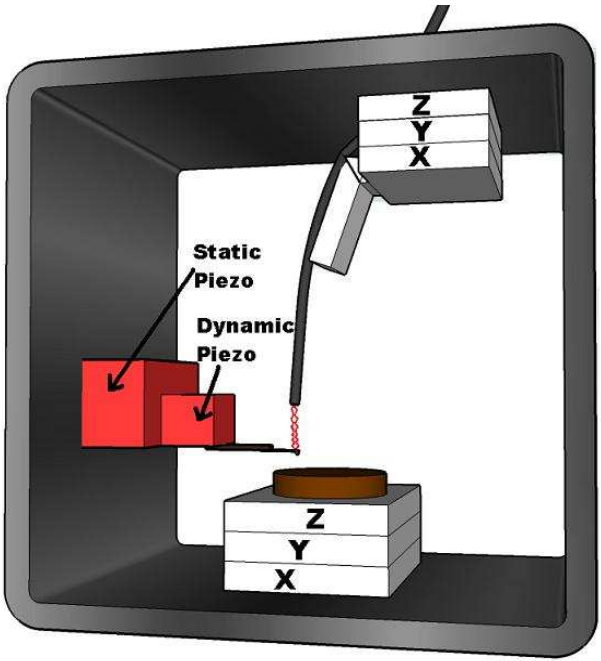}
\caption{The Force Feedback Microscope. A fiber optic based interferometer measures the tip position. A small piezoelectric element displaces the cantilever base to keep contant the tip position ensuring the stability of the static loop. The force gradient and the damping coefficient are measured employing a second dynamic loop imposing a sub-nanometric oscillation amplitude to the tip. The operational mode of the second loop is described elsewere \cite{io12}.}\label{fig:figA} 
\end{center}
\end{figure}

\begin{figure}
\begin{center}
\includegraphics[width=8.5cm]{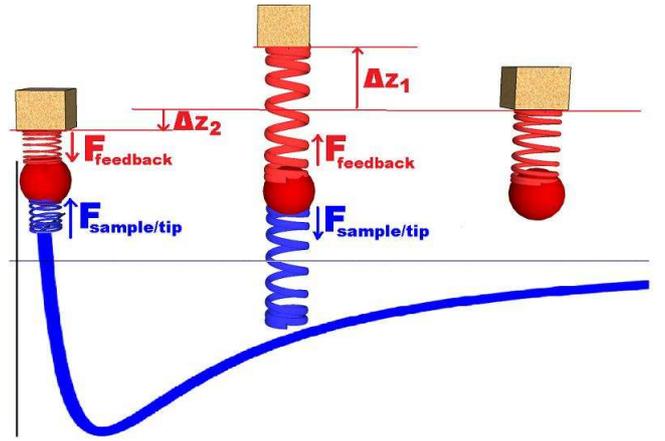}
\caption{The Force Feedback Microscope operational scheme. In presence of the attraction between the tip and the sample, the piezoelectric element displaces the cantilever base up in order to compensate the force acting on the tip. The tip position is therefore kept stable. When the tip-surface force is repulsive, the cantilever base is displaced down.}\label{fig:figB} 
\end{center}
\end{figure}

In the FFM, a fiber optic based interferometer is used to measure the tip position which is kept constant by a static feedback loop (fig.  \ref{fig:figA}).
A second dynamic feedback loop (fig. 1) imposes a constant sub-nanometric oscillation amplitude on the tip. From the measurement of the phase of the tip oscillations and the normalized excitation amplitude as a function of the tip-sample distance, the conservative force gradient and the damping coefficient can be measured \cite{io12}.
In this letter we address the capabilities of the solely static feedback loop. No oscillation amplitude is here imposed to the tip.
The loop ensures that at any time the total force acting on the tip is equal to zero and consequently the tip position is stable.
For this purpose, a small piezoelectric element displaces the cantilever base in order to apply a counteracting force, $F_{feedback} $, on the tip such that: 
\begin{equation}\label{stability}
\sum_{i=1}^2 F_i = F_{sample/tip} + F_{feedback} = 0
\end{equation}
$F_{feedback} $ is equal to the cantilever stiffness $k$ times the displacement $\Delta z$ imposed to the cantilever base by the piezoelectric element.
\begin{equation}
F_{feedback} = k \Delta z
\end{equation}
In fig. \ref{fig:figB}, the operational scheme is briefly presented. In red, the cantilever base is displaced by $\Delta z_1$ in presence of an attractive $F_{sample/tip}$ such that $k \Delta z_1 = F_{sample/tip}$. 
When the surface is further approached to the tip, the interaction $F_{sample/tip}$ gets repulsive and the cantilever base is shifted down by $\Delta z_2$ to compensate the repulsion.
The tip position, in fig. \ref{fig:figB}  represented by the red sphere, is constant over the entire approach curve.
The counteracting force $F_{feedback} $ is therefore a direct measurement of the tip-sample interaction both in the attractive and repulsive regime.\\
In order to fulfill eq.\eqref{stability}, the static loop must control applications of strong feedback forces on the tip over a large bandwidth. In the present work, a proportional-integral-derivative control has been employed to drive the piezoelectric element positioned at the cantilever base. The stability of the feedback loop and the theoretical performances of the instrument are not subjects of this letter and will be discussed in an additional work.
Here we focus on the experimental performances of the static feedback loop.\\
In analogy with the FFM mode \eqref{stability}, in static conventional AFM mode the total force acting on the tip is
\begin{equation}\label{stability2}
\sum_{i=1}^2 F_i = F_{sample/tip} + k  \delta z
\end{equation}
where $\delta z$ is the deflection of the cantilever that is not kept under control during the approach curve.
The jump to contact mechanism occurs when the tip is in a position of unstable equilibrium, therefore when the cantilever stiffness is
\begin{equation}
k = \frac{\partial}{\partial z}(F_{sample/tip})
\end{equation}
As a consequence, part of the attractive interaction is not accessible in static mode because the tip jumps on the next stable equilibrium position where $\sum_{i=1}^2 F_i = F_{sample/tip} + k  \delta z = 0$.\\
It follows that the role of the FFM static loop is to keep the tip in mechanical equilibrium at sample-tip distances where the conventional AFM static mode cannot maintain the tip.
\section{\label{sec:level3}Experimental section:}
\begin{figure}
\begin{center}
\includegraphics[width=8.5cm]{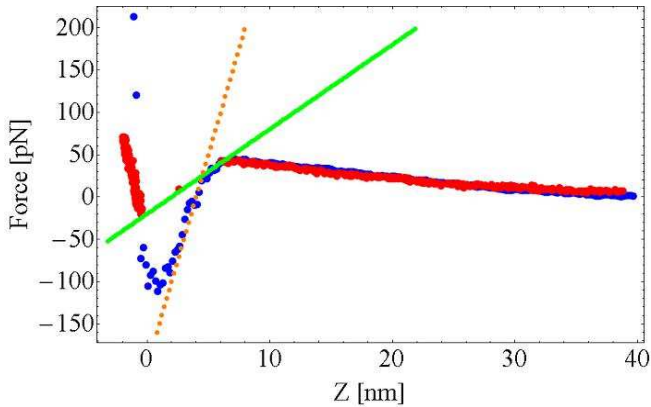}
\caption{Interaction force between mica and a silicon nitride tip in deionized water. The jump to contact occurs when the stiffness of the lever (green line in the graph) is equal to the stiffness of the attractive interaction: here in presence of a \textit{van der Waals} attraction. The stiffness of the cantilever is equal to $0.01 \frac{N}{m}$. Blue) FFM force curve. Red) AFM conventional static force curve. Green) slope of this green line is the cantilever stiffness. Dashed orange) slope is 5 times larger than the cantilever stiffness.}\label{fig:fig1} 
\end{center}
\end{figure}

A DLVO force at the solid/liquid interface is the first example we address.\\
In liquids the short-range Van der Waals interaction between two bodies is usually smaller than in air because of the lower Hamaker constants \cite{bergstrom97} but additional forces could be present and dominate at longer range.
It is the case of the double layer effect at the mica/deionized water interface.
When the electric double layer and the Van der Waals act together, the total interaction is called DLVO, named after Derjaguin, Landau, Verwey and Overbeek \cite{israelachvili07}.
A silicon nitride AFM tip experiences a long range repulsive force and a short-range attractive van der Waals interaction before of the mechanical contact with the mica.\\
A surface of mica glued on a teflon disk has been freshly cleaved with tape. The surface has then been rinsed with $500 \mu l$ of deionized water. A silicon nitride cantilever with a nominal stiffness of $0.01 \frac{N}{m}$ has been used as force transducer.\\
In figure \ref{fig:fig1} the static conventional AFM force curve in red is compared to the FFM force curve in blue.
While conventional AFM static mode is able to measure the double layer contribution to the force, it cannot measure the attractive part of the interaction due to the mechanical instability of the tip occurring in presence of the short-range attractive forces.
The green curve represents the force acting on the tip due to the cantilever stiffness.
The intersection of the green curve and the red curve indicates the tip stable positions before and after the jump to contact \eqref{stability2}. 
The sample/tip force (in blue) is instead fully measured by the Force Feedback Microscope including the van der Waals and chemical contributions dominating at tip-sample distance lower than 5 $nm$.
Situations where tip sample force gradients are five times larger than the cantilever spring constant are here faced by the FFM static feedback loop.
The force curve in the FFM mode has been recorded with an approach speed of 1 $nm/s$.
\\
\begin{figure}
\includegraphics[width=8.5cm]{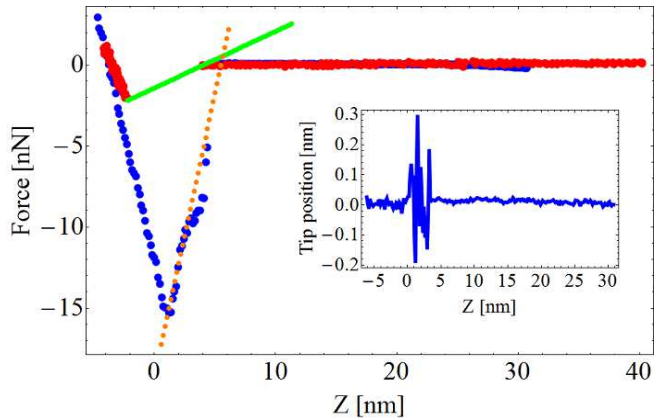}
\caption{\label{fig:fig2} Interaction force between an oxide native silicon surface and a silicon tip in ambient conditions.
The jump to contact occurs when the stiffness of the lever (green line in the graph) is equal to the stiffness of the attractive interaction: in the present case when capillary forces appear. Here the stiffness of the cantilever is equal to $0.35 \frac{N}{m}$. Blue) FFM force curve. Red) AFM conventional static force curve. Green) slope of this green line is the cantilever stiffness. Dashed orange) slope is 10 times larger than the cantilever stiffness.}
\end{figure}
The second example is the capillary condensation at the solid/air interface in ambient conditions.
It has been observed that capillary condensation is an ubiquitous phenomenon at the nanoscale \cite{tabor50,charlaix2010capillary} and it is therefore extremely important because it drives many physical phenomenons such as friction. Moreover, the properties of fluids and confined fluids at the nanoscale revealed to be particularly different than the ones at larger scale \cite{frenken08}. Capillary condensates are therefore of great importance in the nanofluid domain. Capillarity induces a large attractive force between the AFM tip and the sample and therefore it usually dominates in the attractive regime over all others interactions. Since the nucleation time \cite{riedo05} is in the order of the $ms$, the measurement of the interaction is a challenge for the Force Feedback Microscope.\\
Although AFMs techniques are widely used to measure properties of capillary bridges \cite{riedo06,frenken08,barcons12}, the measurement of the complete tip-sample force is still a challenge.
In static mode it could be accessible just employing stiff cantilevers, therefore decreasing the instrumental sensitivity. This leads to a higher incertitude for positioning the tip at a fixed tip-sample distance.
Frequency modulation AFM is not suitable to follow the nucleation force of a capillary condensate because the tip experiences a time dependent interaction and not a reversible force gradient over the oscillation cycle that is concerned by the nucleation.\\
A native oxide silicon surface has been here used as sample.
A cantilever with a nominal stiffness equal to 0.35 $\frac{N}{m}$ with a silicon tip with 10 $nm$ curvature radius is used as the force transducer.\\
In figure \ref{fig:fig2} the static conventional AFM force curve in red is compared to the FFM force curve in blue. The approach speed of the FFM measurement is here 1 $nm/s$.
The mechanical instability occurs here when the capillary condensation leading to a water bridge formation takes place.
In analogy with figure \ref{fig:fig1}, at the intersection of the green line and the conventional AFM force curve we get the stable positions of the tip before and after the jump to contact.
In blue, the sample/tip force is fully measured by the Force Feedback Microscope including the nucleation of the water bridge at a tip-sample distance of 5 $nm$.
The inset gives the stability of the tip position during the approach. It is the error of the static feedback loop.\\
Compared to the force curve acquired at the solid/liquid interface where the counteracting loop just has to ensure the overall stiffness of the force transducer to be positive, at the solid/air interface one additional difficulty has to be overcome. The force due to the nucleation of the water bridge is suddenly established in a short time and therefore the loop must react fast enough. 
Once the capillary bridge is formed, the maximum stiffness that can be compensated is higher than in the case of figure \ref{fig:fig1} and reaches the experimental limits of the instrument. 
\section{\label{sec:level4}Performances:}
The proposed set-up is able to counteract routinely attractive force gradients $5 - 10$ times larger than the cantilever spring constant. Moreover, it shows the capability to counteract interaction forces which are time dependent.
For this purpose, the PID gains of the static feedback loop must be properly chosen to increase the overall stiffness of the force transducer (proportional gain) and to counteract time dependent interactions (integral gain).\\
Force gradients $5 - 10$ times larger than the cantilever spring constant  represent the gain in the instrumental force sensitivity compared to the conventional static mode when an equivalent stiffer cantilever has to be employed to overcome the mechanical instability of the tip. At higher tip-sample force gradients the error of the static loop becomes more important and the measured force is consistently averaged. In presence of an even higher tip-sample force gradients, the jump to contact cannot be avoided.\\
\section{\label{sec:level5}Conclusion:}
The interaction force between two objects at the nanoscale is one of the most important curves in science.
This letter shows the experimental capabilities of a new atomic force microscope when this curve has to be measured in the piconewton and in the nanonewton range showing results for two different interfaces.
The ability to overcome the jump to contact opens the path to the investigation of the full interaction curve including its most specific aspects at short range  distances without mechanical contact between the two objects.
Interaction forces, including capillary condensation, that are not fully accessible directly or indirectly by any other AFM technique force are now measurable.
The static operational scheme could be exported to any other scientific instrument using a spring as a force transducer, such as a Surface Force Apparatus.\\

\noindent \textbf{AKNOWLEDGMENTS} \\
Luca Costa acknowledges COST Action TD 1002.
Mario S. Rodrigues acknowledges financial support from Funda\c{c}\~{a}o para a Ci\^encia e Tecnologia SFRH/BPD/69201/2010. 

\bibliographystyle{ieeetr}
\bibliography{papers}

\end{document}